\documentclass[twocolumn,notitlepage,prb,color,superscriptaddress,psfig,showkeys,showpacs,amsmath,amssymb,nobibnotes,longbibliography]{revtex4-2}

\setcitestyle{square,numbers}

\usepackage{graphicx}
\usepackage[caption=false]{subfig}
\usepackage{float}
\usepackage{epstopdf}
\usepackage{dcolumn}
\usepackage{hyperref}
\hypersetup{
    colorlinks   = true, 
    urlcolor     = blue, 
    linkcolor    = blue, 
    citecolor    = blue 
}
\usepackage{xcolor}
\usepackage{xspace}
\xspaceaddexceptions{]\}}
\usepackage{soul}
\setstcolor{red}
\usepackage{natbib}
\usepackage{amsmath}

\newcommand{\CGT}{Cr$_2$Ge$_2$Te$_6$\xspace}

\newcommand{\vk}[1]{{\color{black}{#1}}}

\begin{document}
	

\title{Pressure control of the magnetic anisotropy \\ of the quasi-two-dimensional van der Waals ferromagnet \CGT}

\author{T.~Sakurai}
\thanks{These authors contributed equally to this work.}
\affiliation{Research Facility Center for Science and Technology, Kobe University, Kobe 657-8501, Japan}
\author{B.~Rubrecht}
\thanks{These authors contributed equally to this work.}
\affiliation{Leibniz IFW Dresden, D-01069 Dresden, Germany}
\affiliation{Institute for Solid State and Materials Physics, TU Dresden, D-01062 Dresden, Germany}
\author{L.~T.~Corredor}
\thanks{These authors contributed equally to this work.}
\affiliation{Leibniz IFW Dresden, D-01069 Dresden, Germany}
\author{R.~Takehara}
\affiliation{Graduate School of Science, Kobe University, Kobe 657-8501, Japan}
\author{M.~Yasutani}
\affiliation{Graduate School of Science, Kobe University, Kobe 657-8501, Japan}
\author{J.~Zeisner}
\affiliation{Leibniz IFW Dresden, D-01069 Dresden, Germany}
\affiliation{Institute for Solid State and Materials Physics, TU Dresden, D-01062 Dresden, Germany}
\author{A.~Alfonsov}
\affiliation{Leibniz IFW Dresden, D-01069 Dresden, Germany}
\author{S.~Selter}
\affiliation{Leibniz IFW Dresden, D-01069 Dresden, Germany}
\affiliation{Institute for Solid State and Materials Physics, TU Dresden, D-01062 Dresden, Germany}
\author{S.~Aswartham}
\affiliation{Leibniz IFW Dresden, D-01069 Dresden, Germany}
\author{A.~U.~B.~Wolter}
\affiliation{Leibniz IFW Dresden, D-01069 Dresden, Germany}
\author{B.~B\"uchner}
\affiliation{Leibniz IFW Dresden, D-01069 Dresden, Germany}
\affiliation{Institute for Solid State and Materials Physics and W{\"u}rzburg-Dresden Cluster of Excellence ct.qmat, TU Dresden, D-01062 Dresden, Germany}
\author{H.~Ohta}
\affiliation{Molecular Photoscience Research Center,  Kobe University, Nada, Kobe 657-8501 Japan}
\author{V.~Kataev}
\affiliation{Leibniz IFW Dresden, D-01069 Dresden, Germany}

%
%

\date{\today}

\begin{abstract}

We report the results of the pressure-dependent measurements of the static magnetization and of the ferromagnetic resonance (FMR) of \CGT to address the properties of the ferromagnetic phase of this quasi-two-dimensional van der Waals magnet. The static magnetic data at hydrostatic pressures up to 3.4\,GPa reveal a gradual suppression of ferromagnetism in terms of a reduction of the critical transition temperature, a broadening of the transition width and an increase of the field necessary to fully saturate the magnetization $M_{\rm s}$. The value of $M_{\rm s} \simeq 3\mu_{\rm B}$/Cr remains constant within the error bars up to a pressure of 2.8\,GPa. The anisotropy of the FMR signal continuously diminishes in the studied hydrostatic pressure range up to 2.39\,GPa suggesting a reduction of the easy-axis  type magnetocrystalline  anisotropy energy (MAE). A quantitative analysis of the FMR data gives evidence that up to this pressure the MAE constant $K_{\rm U}$, although getting significantly smaller, still remains finite and positive, i.e. of the easy-axis type. Therefore, a recently discussed possibility of switching the sign of the magnetocrystalline anisotropy in \CGT could only be expected at still higher pressures, if possible at all due to the observed weakening of the ferromagnetism under pressure. This circumstance may be of relevance for the design of strain-engineered functional heterostructures containing layers of \CGT.  

\end{abstract}

\keywords{van der Waals magnet, 2D material, magnetic anisotropy, pressure effects, ferromagnetic resonance, magnetization}

\maketitle

\section{Introduction}

The layered van der Waals magnet \CGT \cite{Carteaux1995} belongs to a class of materials which are currently the subject of intense research activities as these systems offer, on the one hand, a materials base for experimentally exploring details of two-dimensional magnetism \cite{Park2016,Gong2017,Burch2018}. On the other hand, the weak couplings between individual layers render these materials promising as (magnetic) components in so-called van der Waals heterostructures \cite{Gibertini2019}. In both respects it is essential to properly characterize the magnetic anisotropies of the investigated materials. At zero external pressure, \CGT features a uniaxial magnetic anisotropy with the magnetic easy axis being oriented parallel to the crystallographic $c$ axis, i.e., perpendicular to the honeycomb layers extending in the $ab$ plane (see, e.g., Refs.~\cite{Carteaux1995,Zhang2016,Lin2017,Liu2017}). In a previous study \cite{Zeisner2019} we established the value of the uniaxial magnetocrystalline anisotropy energy density (MAE) $K_{\rm U} = 0.48\times 10^6$\,erg/cm$^3$ by means of ferromagnetic resonance (FMR) measurements under standard conditions at ambient pressure. A detailed discussion of this result and its comparison with the band structure calculations can be found in Ref.~\cite{Zeisner2019}.

One route to modify the magnetic properties of van der Waals magnets is the application of hydrostatic pressure to the crystals, which, in the case of \CGT, was already attempted in Refs.~\cite{Sun2018,Lin2018}. Lin \textit{et al.} \cite{Lin2018} reported a change in the MAE upon applying external pressures up to 2\,GPa, in particular a change in the sign of $K_{\rm U}$. This implies that the anisotropy of the system transforms from \textit{easy-axis type} (easy axis parallel to the crystallographic $c$~axis) to \textit{easy-plane type} (easy direction perpendicular to the $c$~axis and thus in the $ab$~plane). However, details of the MAE as a function of applied pressure have not been investigated so far. This motivates a systematic, combined magnetization and FMR study on \CGT at various hydrostatic pressures aiming at a quantification of the pressure dependence of $K_{\rm U}$. To our knowledge, the FMR experiments on \CGT under pressure and the magnetization studies at pressures exceeding 1\,GPa  were not reported so far.   

In the present work we communicate a comprehensive investigation of the effect of the hydrostatic pressure on the static magnetic and FMR properties of single crystals of \CGT at pressures up to $P = 2.39$ and 3.4\,GPa applied in the FMR and magnetization experiments, respectively. With increasing pressure, the magnetic ordering temperature gradually decreases from $T_{\rm c} = 66$\,K at $P = 0$ down to 45\,K at $P = 3.4$\,GPa. Concomitantly, the transition width broadens and the saturation field increases. Remarkably, the saturation magnetization stays practically constant up to $P= 2.8$\,GPa and decreases at a higher pressure. The FMR measurements performed at $T = 4.2$\,K~$\ll T_{\rm c}$ in the frequency range $50 - 260$\,GHz for the two orientations of the applied magnetic field, ${\bf H}\parallel c$ and ${\bf H}\perp c$, evidence that the excitation energy gap characteristic of an easy-axis ferromagnet closes at $P > 2$\,GPa. The data analysis in the frame of a phenomenological theory of FMR reveals that the seemingly isotropic magnetism of the studied crystals manifesting in the gap closure is a result of the compensation of the intrinsic easy-axis magnetocrystalline anisotropy of \CGT and the shape anisotropy of the sample. It follows from our analysis that the MAE constant $K_{\rm U}$ is reduced by a factor of 2 at the highest applied pressure in the FMR experiment but still remains sizable. This suggests a robustness of the easy-axis type ferromagnetism of \CGT to the application of pressure up to $P = 2.39$\,GPa despite a reduction of $T_{\rm c}$. Our findings  motivate further pressure experiments to address the question of whether the sign of the MAE could be changed at still high pressures before the ferromagnetic state would be fully suppressed. From the technological perspective, the results of the present work on the pressure tunability of the magnetic anisotropy of \CGT may be insightful for accessing the use of this material as a magnetic element of  spintronic devices with strained architecture. 

\section{Samples and methods}

Single crystals of \CGT were grown with the self flux technique and thoroughly characterized as described in detail in our previous work in Ref.~\cite{Zeisner2019}. As grown single crystals were thoroughly characterized by powder x-ray diffraction and energy dispersive x-ray spectroscopy, both agree well with the crystal structure in the R\={3} space group as well as with the expected stoichiometry of \CGT ~\cite{Selter2020}.

The bulk magnetization data were measured using a custom-built pressure cell for a commercial Quantum Design superconducting quantum interference device (SQUID) magnetometer (MPMS-XL). Inside the CuBe cell, two opposing cone-shaped ceramic anvils compress a CuBe gasket with a cylindrical hole used as sample chamber \cite{Alizera2009,Schottenhamel2012}. The plateletlike-shaped single-crystalline sample ($m \sim 3.3 \times 10^{-6}$\,g) of \CGT with the $c$~axis normal to the plate  was installed into the gasket hole (diameter \O{} = 0.6~mm and height $h$ = 0.8~mm) and rested on the flat part of the ceramic anvil. Given the longitudinal magnetic field direction in the SQUID magnetometer the ${\bf H} \parallel c$ orientation was easily achieved.  The uniaxial force is applied at ambient temperature, and it is converted into hydrostatic pressure in the sample chamber using Daphne oil 7575 as the pressure-transmitting medium. We checked the pressure value and its homogeneity at low temperature by measuring the pressure-dependent diamagnetic response associated with the superconducting transition of a small Pb manometer inserted in the sample space~\cite{Bireckoven1988}. Such measurements were performed in applied fields of $H = 5$~Oe. In order to ensure a stable pressure throughout all measurements due to the thermal expansion of the cell upon temperature sweeps, a thermal cycling was performed once from ambient temperature to 2~K and back to ambient temperature before the actual data acquisition.

Note that a magnetic background is detected in our DC magnetization measurements arising from the pressure cell itself. Here, we performed a detailed magnetic characterization of the empty pressure cell within the same conditions as the real experiments with the aim of a quantitative and reliable disentanglement of the sample signal from the overall magnetic response. \vk{The background correction was done by subtracting the fitted signal of a gasket without a sample measured under the same conditions, given that the raw signal of the gasket in our setup can be described by a magnetic dipole in our SQUID detection coils.} While at temperatures above $\sim$ 20~K a temperature-independent pressure cell background ensures excellent background subtraction, at low temperatures a strongly temperature dependent magnetization of the CuBe cell limits the resolution of our experiments. Consequently, only background-subtracted data for $T >$ 10~K are shown. Small kinks in the magnetization curves around 40~K were identified as instrumental artifacts. More detailed information about the background of our pressure cell can be found in Ref.~\cite{Prando2016}. Still, the combination of the very small magnetization above the ferromagnetic ordering temperature of our samples with a very small mass 
($ m \sim 3.3 \times 10^{-6}$ g) 
and the large uncertainty of absolute pressure values for $T > 150$ K using Pb as a manometer at low $T$ in our ceramic anvil pressure cell restrains us from a quantitative Curie-Weiss analysis in the high-temperature regime.

\vk{Ferromagnetic resonance  is a resonance response of the total magnetization of the ferromagnetically ordered material exposed to the microwave radiation. The FMR frequency not only is determined  by the strength of the applied  external magnetic field but also sensitively depends on the magnetic anisotropy of the studied sample, as will be explained in Sec.~\ref{subsection:FMR}. The magnetic anisotropy can be accurately quantified by measuring FMR at different excitation frequencies. Therefore, in the present work}
FMR experiments were carried out using a multifrequency electron spin resonance setup equipped with a piston-cylinder pressure cell with a maximum pressure of about 2.5 GPa. Its detailed description can be found in Ref.~\cite{Sakurai2015}. Microwave radiation with frequencies in the range 50 -- 260\,GHz was provided by a set of Gunn diode oscillators and detected by a $^4$He-liquid cooled hot-electron InSb bolometer. Magnetic fields up to 10\,T were generated by a cryogen-free superconducting magnet. Plateletlike single-crystalline samples of \CGT with typical lateral dimensions of $\sim 3-4$\,mm were put inside the pressure cell in a Teflon capsule filled with a pressure-transmitting fluid (Daphne 7373 oil from Idemitsu Kosan Co., Ltd \cite{NoteOil,Murata2008,Murata2016}). 
The advantage of having a well-defined $c$-axis direction of the samples  facilitated their orientation in the FMR pressure cell. For the ${\bf H} \parallel {\bf c}$ field geometry the studied sample was placed  on the flat bottom of the Teflon capsule. For ${\bf H} \perp {\bf c}$ it was firmly fixed with a thin Teflon tape to the side of a small rectangular parallelepiped placed on the bottom of the Teflon capsule.
The inner pistons of the pressure cell made of ZrO$_2$-based ceramics ensured low-loss propagation of the microwaves through the cell. The pressure was calibrated using a superconducting tin-based pressure gauge. All FMR measurements were performed at a temperature $T = 4.2$\,K for two orientations of the sample with respect to the applied magnetic field,  ${\bf H}\parallel c$~axis and ${\bf H}\perp c$~axis.

\section{Experimental results and discussion}

\subsection{Pressure-dependent magnetization study}

\begin{figure}
	\includegraphics[width=\columnwidth]{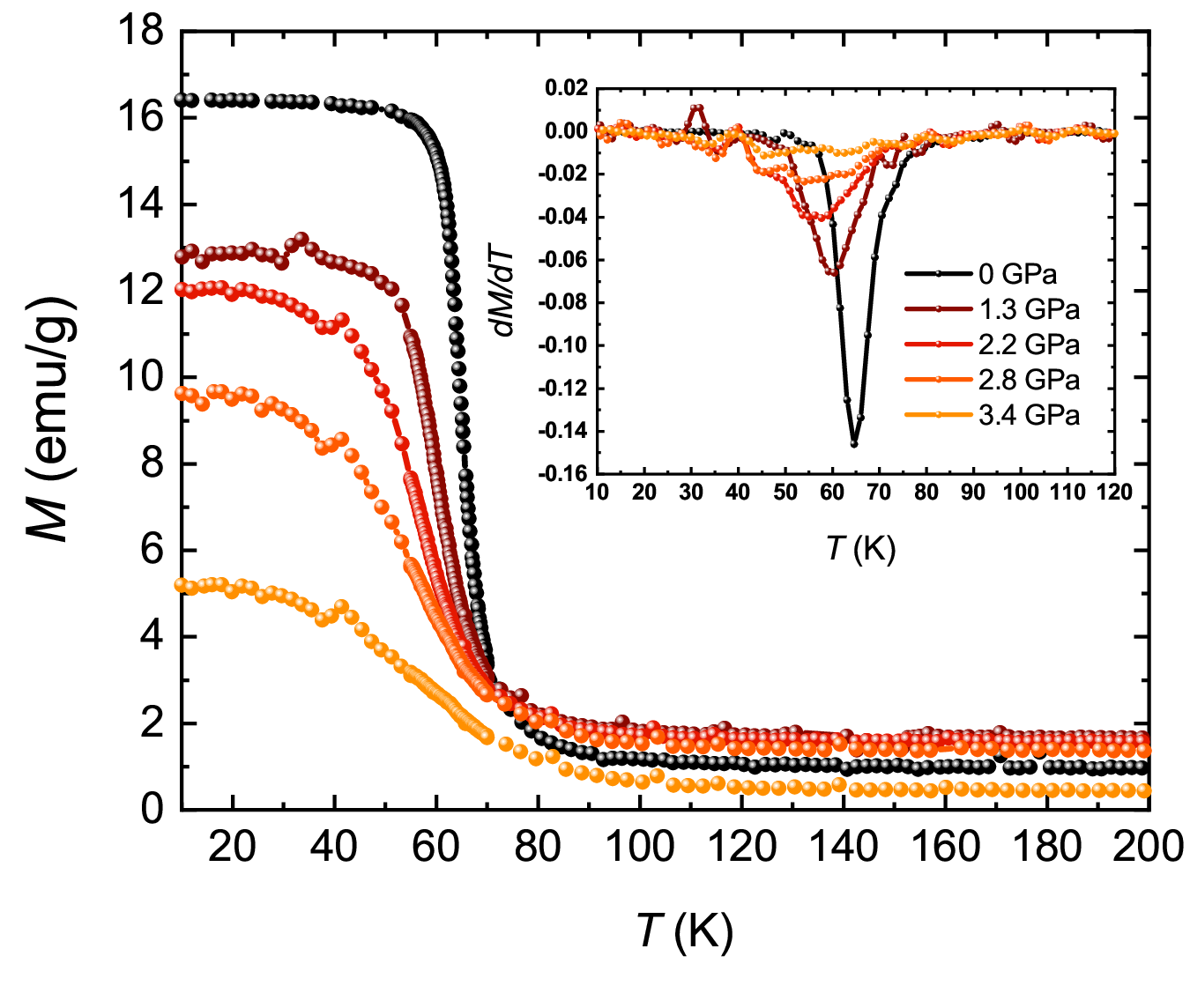}
	\caption{Temperature dependence of the magnetization $M(T)$ of \CGT measured in the FC mode at $\mu_0H = 0.1$\,T parallel to the $c$~axis for pressures up to 3.4\,GPa. The inset shows the corresponding $dM(T)/dT$ curves.}
	\label{fig:MxT}
\end{figure}

\begin{figure}
	\includegraphics[width=\columnwidth]{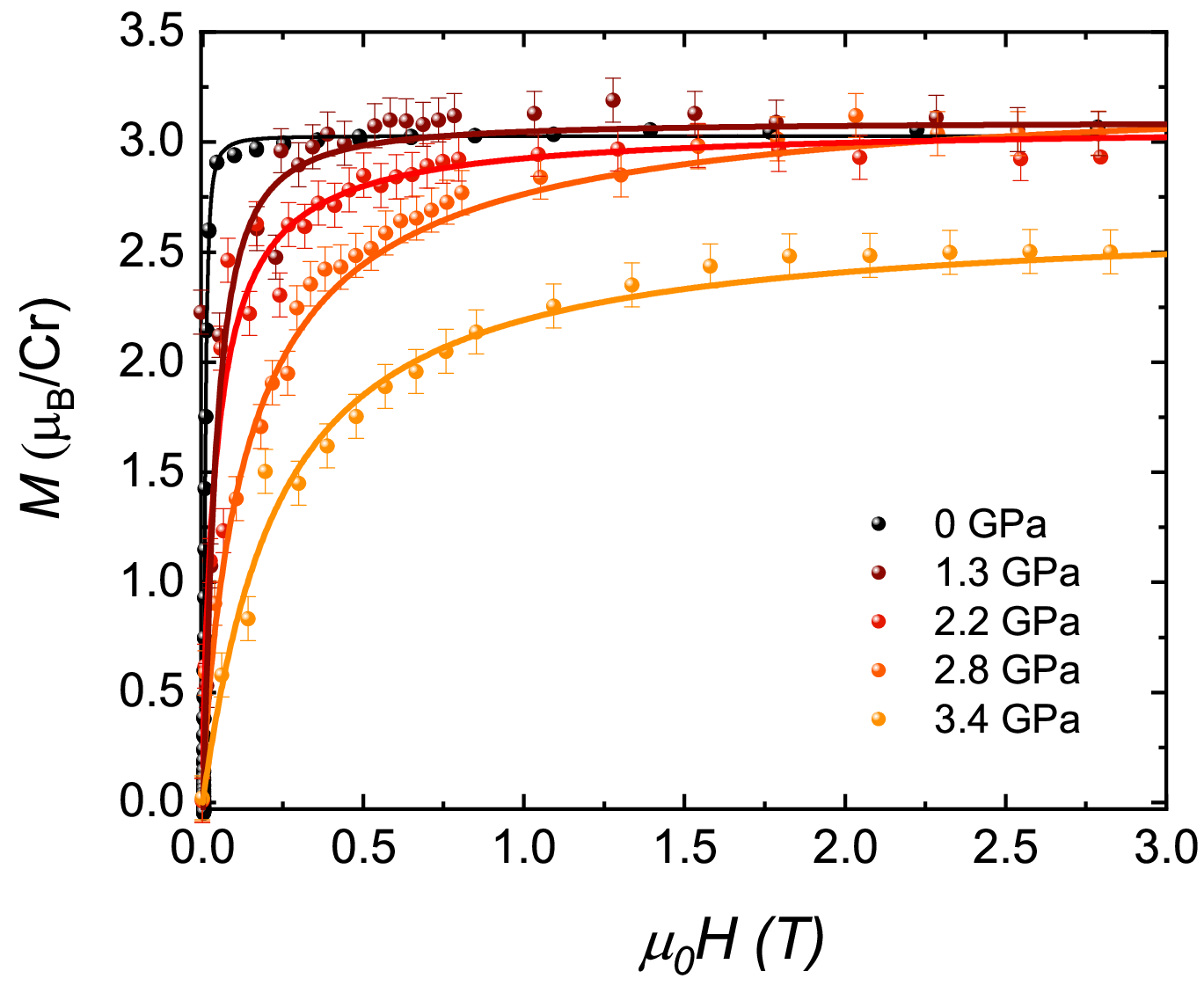}
	\caption{Magnetization $M$ as a function of applied field ${\bf H} \parallel c$ at $T = 15$\,K for pressures up to 3.4\,GPa. The solid lines are guides to the eye.
		 A demagnetization correction has been applied~\cite{Osborn1945}.}
	\label{fig:MxH}
\end{figure}

The field-cooled (FC) magnetization curves $M(T)$ as a function of temperature are shown in Fig.~\ref{fig:MxT} for an applied magnetic field $\mu_0H = 0.1$\,T oriented parallel to the $c$~axis under an applied pressure up to 3.4\,GPa. \vk{Due to the small magnetization values, resulting from the very small mass of the sample, the absolute values at high temperature show a small artificial shift in the magnetization axis, coming from a highly sensitive background subtraction.} Two main characteristics are observed. First, the onset of the magnetic transition at approximately 80\,K is persistent within $\pm 1$\,K despite the pressure change. 
The onset was determined from the derivative $dM/dT$ curve (Fig.~\ref{fig:MxT}, inset) 
while moving from high to low temperature as the point of its departure from a very small, close to zero value $dM/dT \sim 0$ caused by just a small polarization of paramagnetic moments  to the negative values due the gradual development of the spontaneous ferromagnetic moment.
Second, the transition broadens continuously, and the ferromagnetic (FM) transition temperature $T_c$, defined as the minimum of the $dM(T)/dT$ curve \cite{Tcnote,Yeung1986}, shifts to lower temperature upon increasing pressure, as shown in the inset of Fig.~\ref{fig:MxT}. This behavior is in line with an earlier report \cite{Sun2018} in which magnetization measurements at pressures up to 1\,GPa showed the same tendency. Remarkably, the absolute value of the magnetization in the FM state at the given field of the measurement is gradually reduced with increasing pressure up to 3.4\,GPa. This can be explained by a shift of the saturation field $H_{\rm sat}$ as a function of pressure to higher values, which reduces consequently the magnetization at small applied fields such as 0.1\,T (see Fig.~\ref{fig:MxH} and discussion below). 

The magnetic field dependence of the magnetization $M(H)$ is depicted in Fig.~\ref{fig:MxH} for the same pressure values as shown for the temperature dependences. To minimize the influence of the nonperfect subtraction of the background signal from the pressure cell at low temperatures, the curves were measured at $T = 15$\,K, which is still low enough to capture the characteristic behavior of the FM state (see Fig.~\ref{fig:MxT}). The $M(H)$ data set in Fig.~\ref{fig:MxH} includes a reference measurement performed at ambient pressure on a more massive single crystal ($m \sim 4$\,mg) and without the pressure cell. The ambient pressure data yield a saturation moment $M_{\rm s}$ of 3.05\,$\mu_{\rm B}$/Cr, which is consistent with the value found in our previous study~\cite{Zeisner2019}. Importantly, the data in Fig.~\ref{fig:MxH} show that $M_{\rm s}$ is pressure independent up to about 2.8\,GPa within the experimental error bar, the latter mostly being determined by the comparably strong background signal due to the reduced mass of our sample. At higher pressure, however, $M_{\rm s}$ reduces to about 2.5\,$\mu_{\rm B}$/Cr at 3.4\,GPa. Another important observation is that, contrary to the saturation moment, the saturation field $H_{\rm sat}$ continuously shifts to higher values under applied hydrostatic pressure. While saturation can be achieved at $\sim0.1$\,T for ambient pressure, $\mu_0H_{\rm sat}$ increases to $\sim 1.5$\,T at 3.4\,GPa. A possible explanation for the shift of the saturation field  was proposed by Sun {\it et al.}~\cite{Sun2018} through first principles calculations. According to them, the pressure-induced decrease in the Cr-Cr bond length favors antiferromagnetic exchange, while a concomitant deviation from the 90$^\circ$  Cr-Te-Cr bond angle leads to a suppression of the FM superexchange interaction. Increasing  competition between the easy-axis MAE and the easy-plane shape anisotropy under pressure (see  below) may also contribute to an increase of $H_{\rm sat}$. Magnetization measurements at still higher pressures would be enlightening in this regard.

\begin{figure}
	\includegraphics[width=\columnwidth]{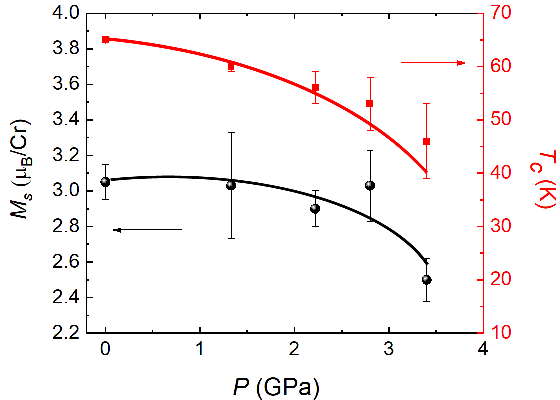}
	\caption{Saturation magnetization $M_{\rm s}$ and the FM transition temperature $T_{\rm c}$ as a function of the applied pressure (symbols). \vk{$M_{\rm s}$ was calculated as the average magnetization value for applied fields $\mu_0H \geq 2$\,T, 
			with 2\,T determined as the lower limit of the applied field
			 at which $dM/dH = 0$ for all the curves within the experimental error bar.} The solid lines are guides to the eye.}
	\label{fig:Ms+Tc}
\end{figure}

The results of the above analysis of the $M(T)$ and $M(H)$ dependences are summarized in Fig.~\ref{fig:Ms+Tc}, where we plot the $M_{\rm s}$ and $T_{\rm c}$ values as a function of the applied pressure with the corresponding experimental error bars. As already noticed above, the ordering temperature value estimated from the first derivative criterion is less accurate for higher pressures due to the broadening of the transition under pressure (Fig.~\ref{fig:MxT}, inset). Still, we could clearly validate the long-range ordered ferromagnetic ground state of \CGT at 4.2\,K and up to 3.4\,GPa, i.e., for the parameter ranges where FMR experiments  were performed (see below). In contrast to $T_{\rm c}$, the FM saturation moment is approximately constant for pressures up to 2.8\,GPa with a subsequent reduction for $P > 2.8$\,GPa.
%
%

\subsection{Pressure-dependent FMR study}
\label{subsection:FMR}

\begin{figure}[t]
	\includegraphics[width=0.7\columnwidth]{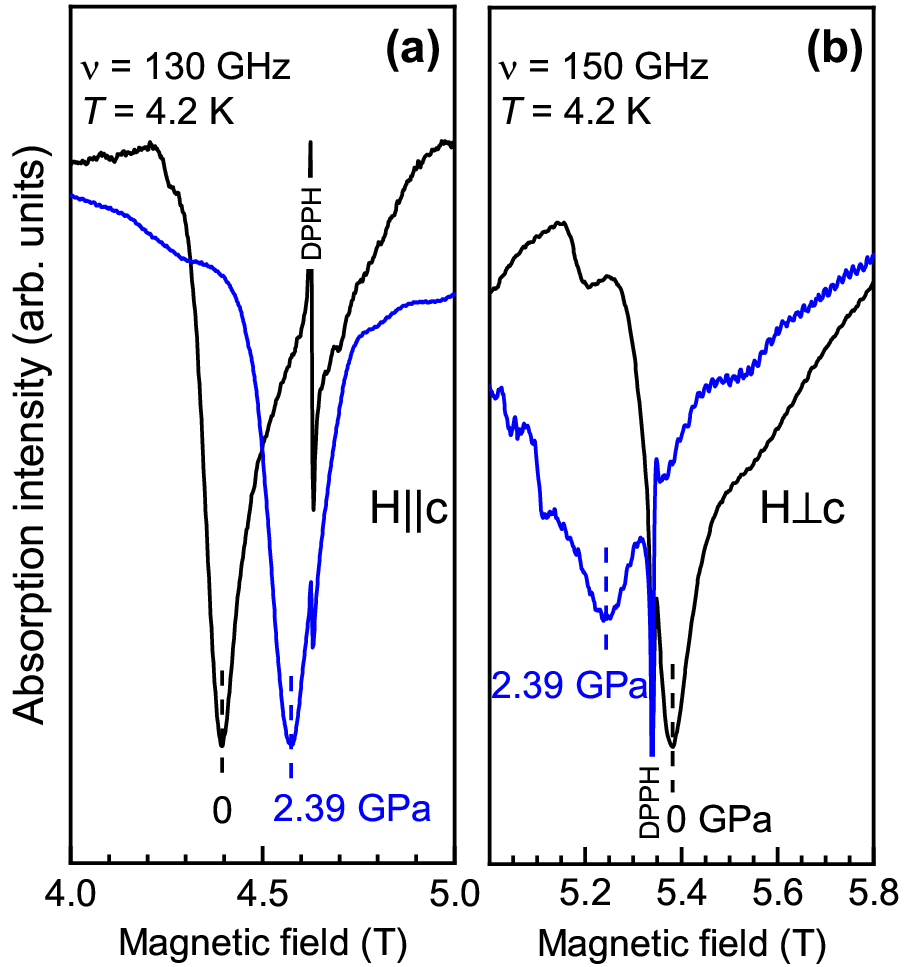}
	\includegraphics[width=0.9\columnwidth]{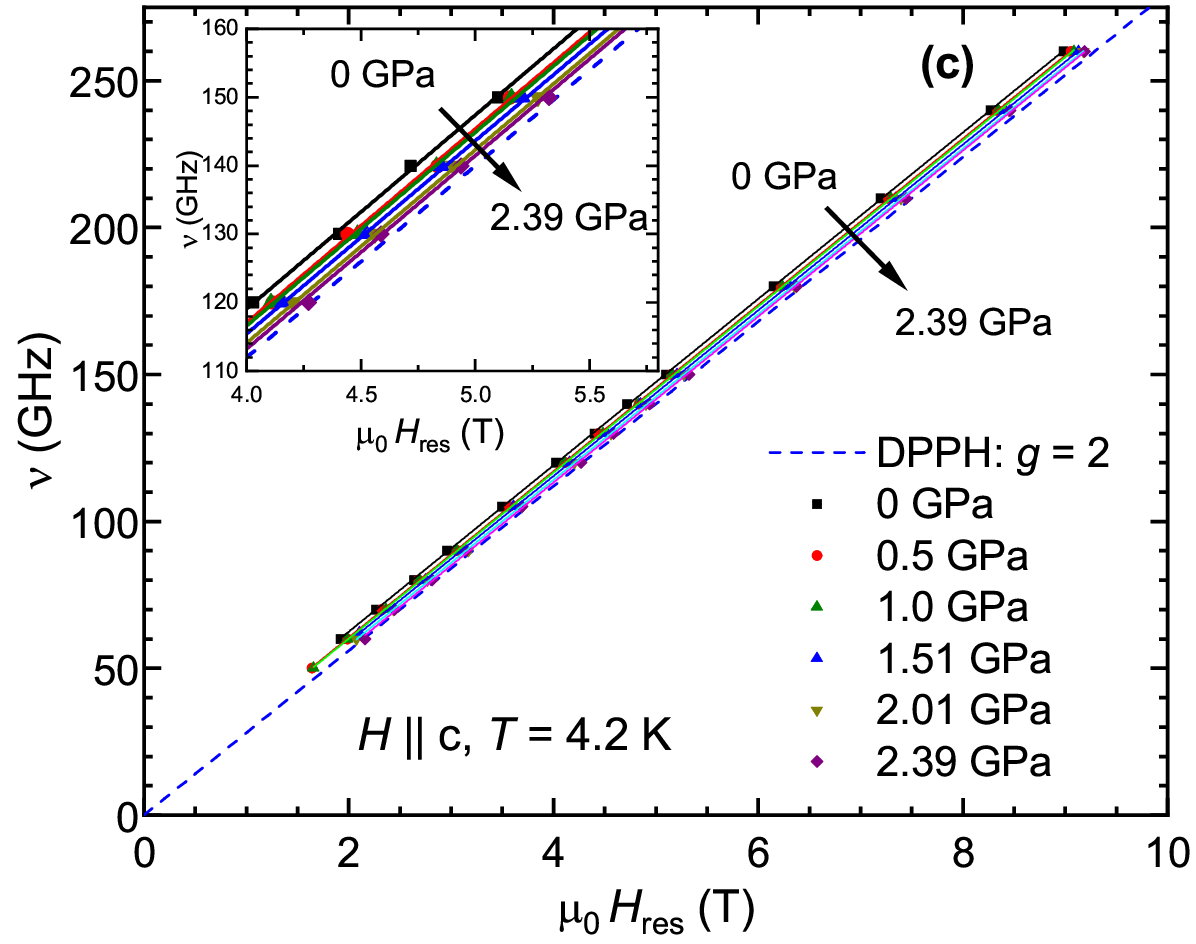}
	\caption{Representative FMR spectra of \CGT at zero and the maximum applied pressure for the applied magnetic field (a) parallel and (b) perpendicular to the $c$~axis. The sharp spike is a signal from the paramagnetic reference DPPH with a $g$~factor of $g\simeq 2$. Some distortions of the lineshape can be ascribed to the interference effects of the microwaves in the pressure cell \cite{notedistort,Eilers94}. (c) The frequency-field diagram obtained from FMR measurements at various pressures up to 2.39\,GPa for ${\bf H}\parallel c$. Solid lines are fits according to Eq.~(\ref{eq:fit_f-dep}) while the dashed blue line indicates the paramagnetic $\nu(H_{\text{res}})$ dependence of the DPPH reference sample with $g\simeq 2$. Details of the frequency-field diagram for intermediate fields are shown in the inset in order to emphasize the systematic shift of the resonance line with increasing pressure.}
	\label{fig:f-dependence}
\end{figure}

The FMR data sets presented in this section were collected by measuring three different single-crystalline samples of \CGT from the same batch with similar, but slightly different, lateral dimensions. As will be discussed below, the differences in demagnetization factors associated with different sample dimensions are, however, negligible in the analysis of the FMR data.

\begin{figure}
	\includegraphics[width=\columnwidth]{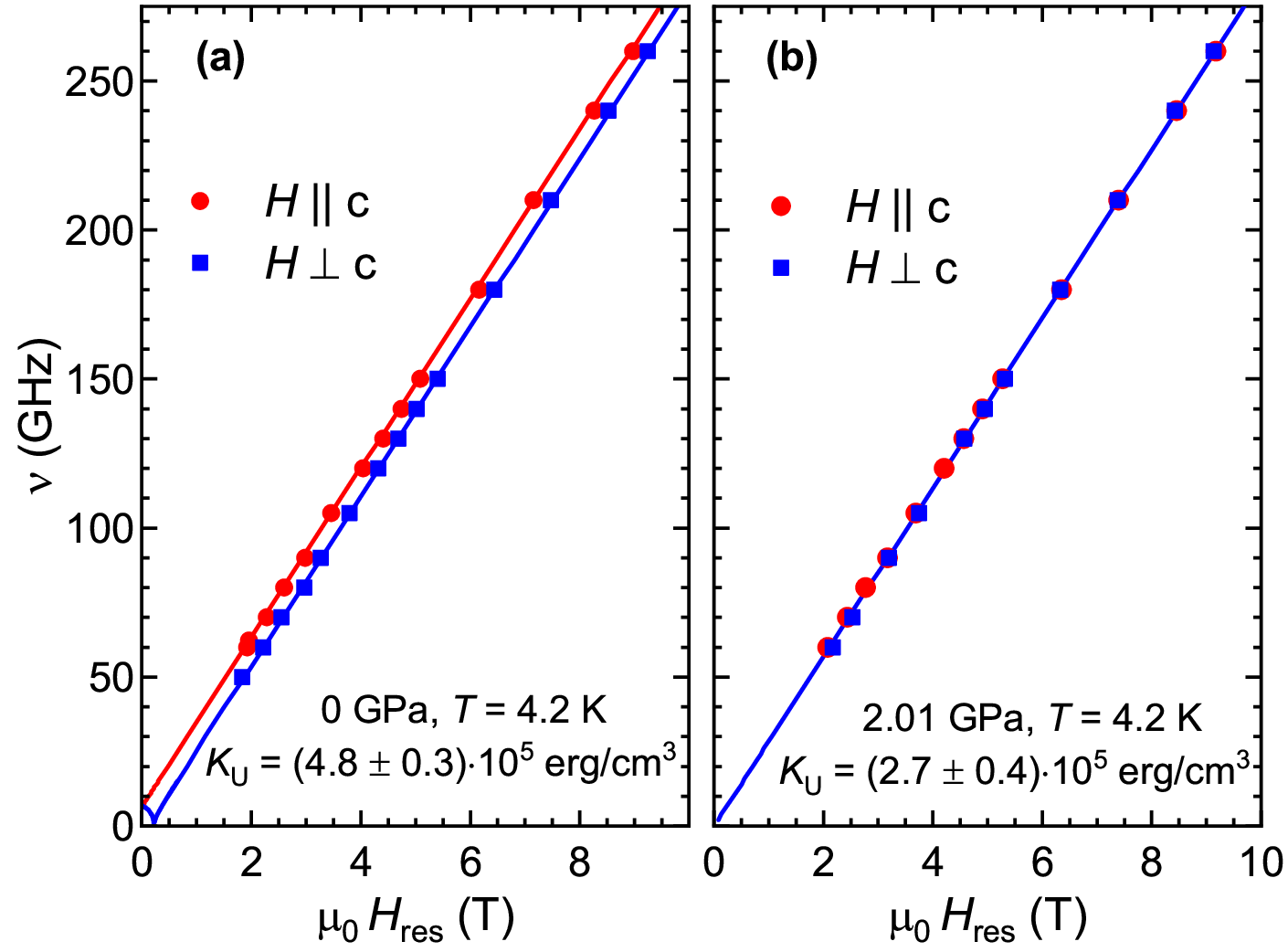}
	\caption{Examples of simulations (solid lines) of the measured frequency-field dependences (symbols) at \mbox{(a) 0\,GPa} and \mbox{(b) 2.01\,GPa}. In these simulations experimental data sets obtained for ${\bf H} \parallel c$ (circles) and ${\bf H} \perp c$ (squares) were taken into account simultaneously in order to determine the value of $K_{\rm U}$.}
	\label{fig:f-dependence_0GPa}
\end{figure}

Selected FMR spectra measured for the two field geometries used in this work are shown in Figs.~\ref{fig:f-dependence}(a) and (b). A shift of the signal to higher fields for ${\bf H}\parallel c$ and to smaller fields for  ${\bf H}\perp c$ by applying pressure is clearly visible.
The results of frequency-dependent FMR measurements at a temperature of 4.2\,K with ${\bf H} \parallel c$ and at different applied external pressures between 0 and 2.39\,GPa are summarized in Fig.~\ref{fig:f-dependence}(c). At 0\,GPa, \CGT features an easy-axis-type MAE with the easy axis oriented parallel to the $c$ axis as was investigated in detail in our previous study \cite{Zeisner2019}. This kind of anisotropy directly manifests in an FMR measurement by a shift of the resonance line towards smaller magnetic fields with respect to the resonance position in the paramagnetic state if the external magnetic field ${\bf H}$ is applied parallel to the magnetic easy axis ($c$~axis) and towards higher magnetic fields if ${\bf H}$ is normal to the magnetic easy axis.
This behavior is illustrated in Figs.~\ref{fig:f-dependence}(a) and \ref{fig:f-dependence}(b), where, depending on the direction of the applied field, the FMR signal at $P = 0$ is shifted either to the left or to the right side of the signal from the paramagnetic reference sample 2,2-diphenyl-1-picrylhydrazyl (DPPH) with a $g$ factor $g\simeq 2$, which is commonly used as a magnetic field marker. Remarkably, the application of an external pressure counteracts these tendencies by shifting the FMR line towards the paramagnetic position for both field geometries. As a result, with increasing the applied pressure from $P = 0$  to 2.01\,GPa the FMR signal from the sample becomes isotropic; that is, the resonance field $H_{\rm res}$ does not depend on the direction of the applied field at this value of pressure [Fig.~\ref{fig:f-dependence_0GPa}(b)].

The frequency-field dependence of the paramagnetic reference sample DPPH is shown in Fig.~\ref{fig:f-dependence} as a blue dashed line for comparison with the ferromagnetic resonance fields $H_{\text{res}}$ measured on the \CGT samples for ${\bf H}\parallel c$~axis. $H_{\rm res}$ shifts continuously to higher magnetic fields with increasing $P$ and thereby approaches, but does not cross, the paramagnetic resonance branch in the frequency-field diagram. This systematic shift is emphasized in the inset of Fig.~\ref{fig:f-dependence}, where details of the frequency-field diagram are shown for intermediate field strengths. The change in the FMR line position as a function of external pressure evidences a reduction of the MAE of the studied system. However, the shift of the resonance fields of \CGT is determined not only by the strength and the sign of the MAE constant $K_{\rm U}$ but also by demagnetization fields resulting from the specific sample shapes which favor, in the case of the platelike-shaped \CGT crystals, an in-plane magnetization. As a consequence, the internal demagnetization fields will lead to a shift of the resonance position to higher fields if the external magnetic field is applied perpendicular to the honeycomb layers, i.e., for ${\bf H}\parallel c$~axis. Thus, the size of the anisotropy constant $K_{\rm U}$ derived from shifts of the resonance position would be underestimated, if demagnetization effects were not taken into account (see the related discussion in Ref.~\cite{Zeisner2019}). These demagnetization fields $H_{\rm D}$ can be calculated using the elements of the demagnetization tensor $N_i$  ($i = x,y,z$, with $x,y,z$ being the principal axes of the tensor) \cite{Osborn1945,Cronemeyer1991,Blundell2001}:
\begin{equation}
 H_{\rm D}^{\rm i} = -4\pi N_{\rm i}M \ \ \ ,
\end{equation}
where $M$ denotes the magnetization of the sample. Since the thickness of the plate-like samples used in this study could not be quantified precisely, the real sample shape was approximated using the demagnetization factors for a flat (infinite) plate oriented perpendicular to the $z$~axis (which was chosen to coincide with the crystallographic $c$ axis of \CGT), i.e., $N_{\rm x} = N_{\rm y} = 0$, $N_{\rm z} = 1$. This assumption is supported by the fact that the lateral dimensions in the $ab$ plane of the used \CGT crystals were much larger than the thicknesses of the samples parallel to the $c$~axis. Using the value of the saturation magnetization $M_{\rm s} \simeq 3$\,$\mu_{\rm B}$/Cr (which corresponds to $M_{\rm s} \approx 201$\, erg/G\,cm$^3$ for \CGT) yields a demagnetization field $\mu_0 H_{\rm D}$ in the fully saturated ferromagnetic phase of -0.253\,T if the external field is applied parallel to the $c$ axis. The frequency-field dependences measured with this field orientation were then fitted by the following expression:
\begin{equation}
\nu_{\text{res}} = \frac{g\mu_{\rm B}\mu_0}{h}[H_{\text{res}}+H_{\rm D}+H_{\rm A}] \ \ \ ,
\label{eq:fit_f-dep}
\end{equation}
where $\nu_{\text{res}}$ denotes the resonance/microwave frequency and $H_{\rm A}$ is the anisotropy field describing the intrinsic magnetocrystalline anisotropy. Here, $\mu_\text{B}$, $\mu_0$ and $h$ are Bohr magneton, magnetic permeability and Plancks constant, respectively. Fits of Eq.~(\ref{eq:fit_f-dep}) to the measured data are shown in Fig.~\ref{fig:f-dependence}(c) as solid lines. For the fitting, the $g$~factor and  $H_{\rm A}$ were treated as free fit parameters. 
The $g$~factor varied very little within the error bars around the mean value of 2.03 (see, Sec.~\ref{sec:simulations}) whereas $H_{\rm A}$ significantly decreased with increasing pressure.  Using the latter parameter, it is possible to calculate the uniaxial MAE $K_{\rm U}$ according to (see, for instance, Ref.~\cite{Skrotskii1966})
\begin{equation}
K_{\rm U}^{\text{fit}} = \frac{H_{\rm A}M_{\rm S}}{2} \ \ \ .
\end{equation}
The results obtained from this fitting procedure are discussed in the following section together with results from an alternative approach for a quantitative analysis of the FMR data.

\section{Simulations of the frequency dependence and determination of the magnetic anisotropies}
\label{sec:simulations}

In order to simultaneously take into account the results of our FMR measurements performed with ${\bf H} \parallel c$ and ${\bf H} \perp c$ field geometries, the measured frequency-field dependences were numerically simulated for each pressure value. This lead to a refinement of the determination of $K_{\rm U}$, in particular at higher pressures. For the simulations we used a well-established phenomenological model of FMR (see, e.g., Refs.~\cite{Smit1955, Skrotskii1966, Farle1998}) in which the resonance frequency $\nu_{\text{res}}$ is expressed as
%
\begin{equation}
\label{eq:omega_res}
\nu_{res}^2 = \frac{g^2 \mu_B^2}{h^2 M_s^2\sin^2\theta} \bigg(\frac{\partial^2{F}}{\partial \theta ^2}\frac{\partial^2{F}}{\partial \varphi^2} - \Big(\frac{\partial^2{F}}{\partial \theta \partial \varphi}\Big)^2\bigg) \ \ \  .
\end{equation}
Here, $F$ is the free-energy density, and $\theta$ and $\varphi$ are the spherical coordinates of the magnetization vector $\boldsymbol{M}(M_s, \varphi, \theta)$.


This phenomenological model is applicable for ferromagnets with fully saturated magnetization at temperatures sufficiently lower than $T_c$, i.e., the case of the fully developed ferromagnetic phase. According to the above-discussed results of our pressure-dependent magnetization study, both criteria are safely fulfilled for the $T$, $P$, and $H$ parameter ranges of the FMR study.


To obtain the resonance position of the FMR signal, $F$ in Eq.~(\ref{eq:omega_res}) should be taken at the equilibrium angles $\varphi_0$ and $\theta_0$ of $\boldsymbol{M}$. In the simulations, the minimum of $F$ with respect to $\theta$ and $\varphi$ for a given set of experimental parameters, such as the microwave frequency and the direction and the strength of the magnetic field, was found numerically. For \CGT, accounting for both the intrinsic uniaxial magnetic anisotropy and of the (extrinsic) shape anisotropy of the particular studied sample enabled an accurate description of all FMR data sets. In this case the free energy density (in cgs units) is defined as      
\begin{equation}
\label{eq:free_energy_density}
\begin{split}
F &= -\boldsymbol{H}\cdot\boldsymbol{M} - K_U\cos^2(\theta) + \\ & \ \ \ \  2\pi M^2_s(N_x\sin^2(\theta)\cos^2(\varphi) + \\ & \ \ \ \  N_y\sin^2(\theta)\sin^2(\varphi) + N_z\cos^2(\theta)) \ \ \ 
\end{split}
\end{equation}
and comprises, besides the Zeeman energy density expressed by the first term, contributions due to the uniaxial and shape anisotropies, the second and third terms, respectively. 
%

\begin{figure}
	\includegraphics[width=0.9\columnwidth]{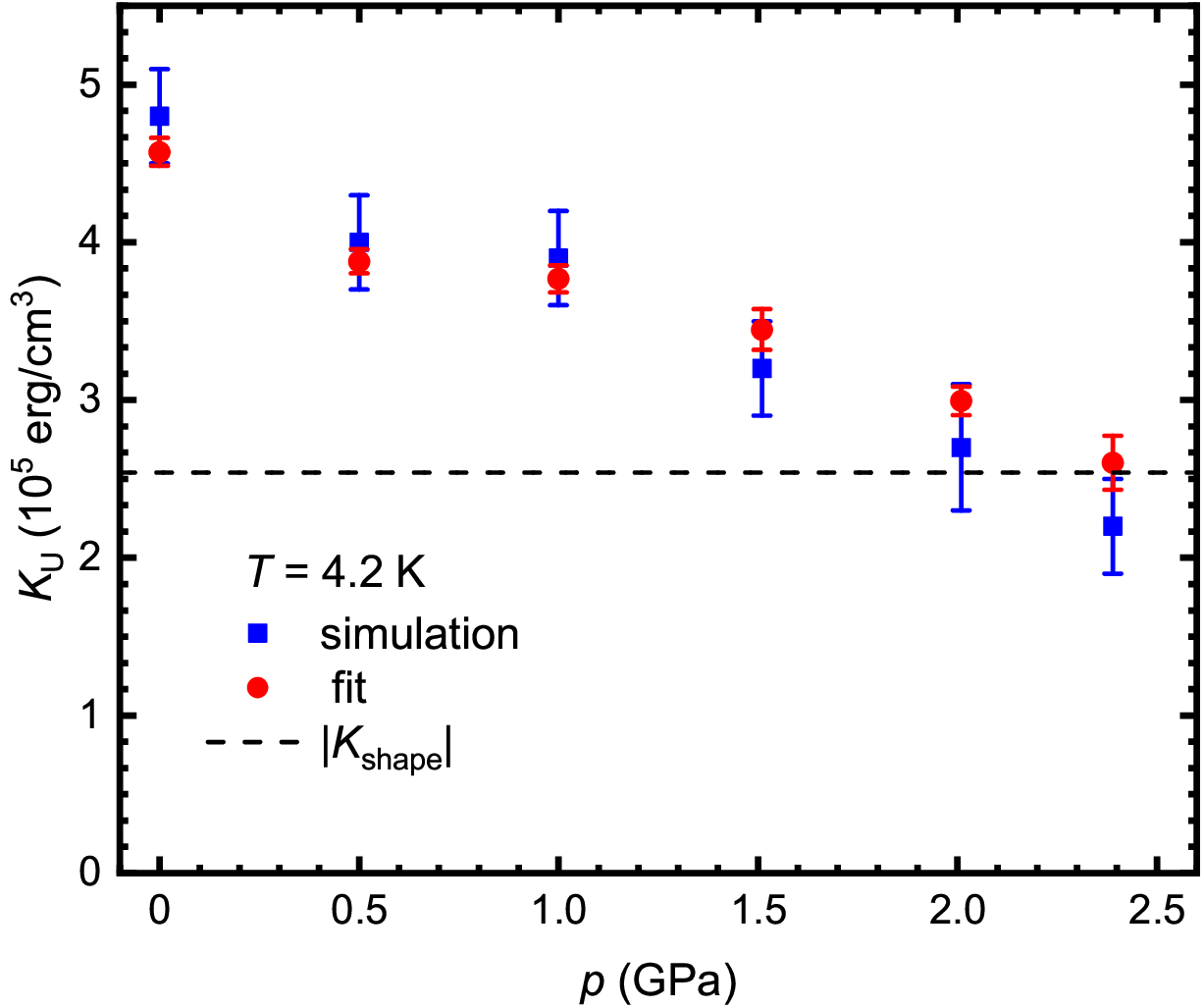}
	\caption{Pressure dependence of the uniaxial magnetocrystalline anisotropy constant $K_{\rm U}$ of \CGT. Red circles denote the results obtained from fitting the different $\nu(H_{\text{res}})$ curves according to Eq.~(\ref{eq:fit_f-dep}) for ${\bf H} \parallel c$. Blue squares are the $K_{\rm U}$ values determined from simulations according to Eq.~(\ref{eq:free_energy_density}) considering the experimental data sets for ${\bf H}\parallel c$ and ${\bf H}\perp c$ simultaneously. \vk{The horizontal dashed line shows the absolute value of the negative shape anisotropy constant $K_{\rm shape} \approx -2.54 \times 10^5$\,erg/cm$^3$. It compensates the decreasing positive value of $K_{\rm U}$ at $p \gtrsim 2$\,GPa yielding a seemingly isotropic behavior.}}
	\label{fig:p-dependence_Ku}
\end{figure}

Representative simulations of frequency-dependent measurements at 0 and 2.01\,GPa are shown in Figs.~\ref{fig:f-dependence_0GPa}(a) and \ref{fig:f-dependence_0GPa}(b), respectively. At 0\,GPa, the separation of the $\nu(H_{\text{res}})$ curves obtained for the two different field orientations clearly evidences the easy-axis-type magnetic anisotropy with the ${\bf H} \parallel c$ curve being shifted to smaller fields and the ${\bf H} \perp c$ curve being shifted to higher fields compared to the paramagnetic position. The 0\,GPa data could be simulated using the value of $K_{\rm U}$ of $4.8 \times 10^5$\,erg/cm$^3$ obtained in our previous study \cite{Zeisner2019}. Keeping the value of $K_{\rm U}$ fixed, the $g$~factor and demagnetization factors were adjusted in order to consistently describe different studied samples. It turned out that all 0\,GPa data could be described successfully with an isotropic $g$ factor of 2.03, which is consistent with our previous investigations, and the same set of demagnetization factors $N_x = N_y = 0$, $N_z = 1$ despite the (small) differences in the shapes of the samples. These parameters were kept fixed in the simulations of the frequency dependences measured at several nonzero pressures. The only free parameter in these simulations was the anisotropy constant $K_{\rm U}$ which allowed us to determine the pressure dependence of the MAE. As can be seen in Fig.~\ref{fig:f-dependence_0GPa}(b), at a pressure of 2.01\,GPa, the measured frequency dependences are nearly identical for both orientations of the external magnetic field (a similar frequency-field diagram was obtained for 2.39\,GPa), giving the impression of isotropic behavior. However, the contributions of the magnetocrystalline anisotropy and the shape anisotropy to the free-energy density [Eq.~(\ref{eq:free_energy_density})] have opposite signs within the pressure range of this study. Thus, the apparently isotropic frequency dependence is, in fact, a consequence of the similar strengths of these two contributions. This allows us to conclude that an application of pressures between 2.01 and 2.39\,GPa reduces the uniaxial MAE $K_{\rm U}$ to a value that compensates the shape anisotropy of the crystals. \vk{With the demagnetization field $\mu_0H_{\rm D} = -0.253$\,T and saturation magnetization $M_{\rm s} \approx 201$\, erg/G\,cm$^3$ estimated above, one obtains the shape anisotropy constant $K_{\rm shape} = (H_{\rm D}M_{\rm s})/2 \approx -2.54 \times 10^5$\,erg/cm$^3$. Its absolute value is indicated by the dashed line in Fig.~\ref{fig:p-dependence_Ku}.}
Moreover, the pressure-dependent FMR studies do not provide any evidence for a pressure-induced sign change of the magnetocrystalline anisotropy or a vanishing of the intrinsic uniaxial anisotropy up to pressures of 2.39\,GPa. This is in contrast to the conclusions drawn in Ref.~\cite{Lin2018} on a spin reorientation transition at pressures between 1.0 and 1.5\,GPa  based on magneto-transport investigations of \CGT.

The pressure dependence of $K_{\rm U}$ as obtained from fitting of the ${\bf H} \parallel c$ data to Eq.~(\ref{eq:fit_f-dep}) and from the simulations of all data available for both field orientations according to Eq.~(\ref{eq:free_energy_density}) is shown in Fig.~\ref{fig:p-dependence_Ku}. Qualitatively, both methods reveal a similar behavior, in particular a reduction of $K_{\rm U}$ with increasing pressure. However, the deviation between the $K_{\rm U}$ values derived from the fits and the simulation increases at higher pressures. This could be attributed to the fact that the simulations simultaneously take into account the frequency dependences measured for both field orientations. Therefore, the reliability of the $K_{\rm U}$ values from simulations is higher despite the larger error bars, in particular at higher pressures at which differences between both field orientations become very small. Finally, it is worth mentioning that in Ref.~\cite{Yu2019} a pressure-induced difference in the unit-cell volume of \CGT between applied pressures of 0 and 3\,GPa of about 4\,\% was reported. As the unit-cell volume enters into the calculation of the saturation magnetization, such a reduction of volume with increasing pressure could change the absolute value determined for $K_{\rm U}$. However, it was found that the pressure-induced contraction of the unit-cell volume by 4\,\% does not affect the determination of $K_{\rm U}$ but is within the given error bars.
%

%
%

\section{Conclusions and outlook}

Our combined magnetization and ferromagnetic resonance study on single crystals of the quasi-two-dimensional ferromagnet \CGT has revealed a significant effect of the application of hydrostatic pressure $P$ on the properties of the ferromagnetic state of this compound. It manifests in a reduction of the critical temperature $T_{\rm c}$ and a concomitant broadening of the transition to the FM state, as well as in a gradual increase of the saturation field $H_{\rm sat}$. In contrast, the saturation magnetization remains practically constant up to $P = 2.8$\,GPa and reduces by approaching the highest applied pressure of 3.4\,GPa. The frequency-dependent FMR measurements performed in the fully developed FM state at $T\ll T_{\rm c}$ and $H > H_{\rm sat}$ at various external pressures up to 2.39\,GPa revealed a systematic shift of the resonance fields with increasing $P$ evidencing a continuous reduction of the magnetocrystalline anisotropy constant $K_{\rm U}$. From the quantitative analysis of the data, which takes into account demagnetization effects, it follows that $K_{\rm U}$, although being significantly reduced at the highest pressures applied in this study, neither vanishes nor changes its sign up to pressures of about 2.4\,GPa. Therefore, noting that FMR is a very direct method for the quantitative determination of magnetic anisotropy, it can be concluded with confidence that there is no evidence for a switching of the magnetic anisotropy from the easy-axis to the easy-plane type within the pressure range under consideration. Further pressure-dependent studies on \CGT are desired to understand the apparent discrepancy between our work and the magnetotransport study in Ref.~\cite{Lin2018} claiming a spin-reorientation transition in \CGT at pressures $1 < P < 1.5$\,GPa and to address the question of whether a pressure-induced sign switching of the magnetic anisotropy could be achieved at pressures exceeding 2.39\,GPa.

Although in terms of the spatial dimensionality we investigated bulk, three-dimensional crystals of \CGT, we showed in our previous electron spin resonance/FMR study that the low-temperature magnetism of the bulk crystals is essentially of a two-dimensional (2D) nature due to very weak interlayer magnetic coupling in this van der Waals compound \cite{Zeisner2019}. Considering the intrinsic magnetic two-dimensionality of the bulk material, our pressure dependent study could be relevant for the research on \CGT in the truly 2D spatial limit.    
In this respect, the pressure control of the magnetic anisotropy investigated in our work may provide important hints for a targeted design of functional magneto-electrical heterostructures containing layers of \CGT. 

In particular, a recent prediction of remarkable strain and electric field tunability of a single layer \CGT\ is encouraging \cite{Wang2019}. It remains yet an open question whether the design of a heterostructure with strain $\epsilon$  of the order of $\pm (1 - 2)$\,\% for the \CGT\  layer, as proposed in Ref.~\cite{Wang2019}, can be achieved. However, a sizable effect on the magnetic anisotropy  which we observed at hydrostatic pressures up to 2.39\,GPa corresponds to an even smaller compressive in-plane strain $\epsilon \sim 0.7$\% \cite{Yu2019}. Such straining seems plausible to achieve since, in general, 2D van der Waals materials are known to sustain large strain.	For example, it was shown that biaxial compressive and tensile strain of $\sim 1$\,\% can be achieved in single-layer  molybdenum dichalcogenide deposited on a thermally compressed or expanded polymer substrate \cite{Frisenda2017}.

\section*{Acknowledgments}

We thank Gael Bastien and Randirley Beltran Rodriguez for technical assistance with the pressure cell used for the magnetization studies, and  the IFW workshop and Juliane Scheiter for the production and treatment of the CuBe gaskets.
The work in Kobe was partially supported by Grants-in-Aid for Scientific Research (C) (Grant No.~19K03746) from the Japan Society for the Promotion of Science.
The work in Dresden was supported by the Deutsche Forschungsgemeinschaft (DFG) through Grant No. KA1694/12-1 and within Collaborative Research Center SFB~1143 ``Correlated Magnetism – From Frustration to Topology” (Project No. 247310070) and the Dresden-Würzburg Cluster of Excellence (EXC 2147) ``ct.qmat - Complexity and Topology in Quantum Matter" (Project No. 390858490). S.A. acknowledges financial support from the DFG through Grant No. AS 523/4-1. 
A.A. acknowledges financial support form the DFG through Grant No. AL 1771/4-1.
S.S. acknowledges financial support from the GRK-1621 Graduate Academy. V.K. gratefully acknowledges the hospitality and financial support during his research visit to Kobe University.


\begin{thebibliography}{33}
\bibitem [{not()}]{notedistort}%
	\BibitemOpen
	\href@noop {} {}\bibinfo {note} {To minimize the undesired interference
			effects in the pressure cell and, in particular, to avoid the interference
			inside the sample we followed the recipe suggested in Ref.~\cite{Eilers94}
			and used very thin crystals with a thickness of a few tens $\mu$m which is
			much smaller than the shortest microwave wavelength of 1153\,$\mu$m at the
			highest applied frequency of 260\,GHz. Furthermore, we repeated each
			experiment several times to confirm that the resonance fields agree within
			the error [smaller than the symbol size in Fig. 4(c)]. This ensured that the
			intrinsic resonance position was always correctly obtained regardless the
			distortion.}\BibitemShut {Stop}%
\end{thebibliography}

%

\end{document}